\documentclass[doublespacing]{elsart}
\usepackage{graphicx}
\begin{document}
\begin{frontmatter}
\title{Alkali-metal ion in rare gas clusters: global minima\thanksref{postadr}}
\author{G. Bilalbegovi\'c}
\ead{goranka@email.hinet.hr}
\thanks[postadr]{To be published in Physics Letters A}
\address{Department of Physics, University of Rijeka, Omladinska 14, 51000 Rijeka, Croatia}

\begin {abstract}
Structural optimization for heteronuclear clusters consisting of
one alkali-metal ion and of up to $79$ neutral rare gas  atoms has
been carried out. The basin-hopping Monte Carlo minimization
method of Wales and Doye is used. Rare gas atoms interact with the
Lennard--Jones potential, whereas the interaction between a
neutral atom and an ion impurity is given by the Mason--Schamp
potential. Starting from eight rare gas atoms the alkali-metal ion
is always inside a cluster.
\end {abstract}
\begin{keyword}
Global optimization \sep Basin-hopping method \sep
Nanoparticles
\PACS 61.46.+w \sep36.40.Mr\sep 36.40.Qv
\end{keyword}
\end{frontmatter}

\section{Introduction}

Rare gas clusters are important and often studied aggregates of
atoms \cite{Echt}. Their mass spectra show the presence of magic
numbers which correspond to the most stable geometrical structures
\cite{Harris}. Because of the
simple form of van der Waals chemical bonding
rare gas clusters are very suitable for theoretical studies. For each
cluster size different isomeric structures exist. They are local
minima in the multidimensional potential energy surface.
The number of these isomers increases rapidly with the size of a cluster.
Several methods of a search for global minima were invented \cite
{Xiao,Deaven,Wolf,Zeiri,Wales,Doye,Leary,Naumkin1,Naumkin2,Miller,Munro,CCD}.
When the size of clusters increases it becomes more difficult to
find the lowest minimum on the potential energy surface.
Optimization methods locate only the candidate for a global minimum.

The problem of interaction between various clusters and impurities
generated intensive research
\cite{Toennies,Hiura,Kumar,Perera,Etters,Velegrakis}. For example,
impurities in helium nanoparticles were studied \cite{Toennies}.
One goal was to determine whether impurity ions, atoms, or
molecules preferably reside in the interior of the cluster, or on
its surface. In general, the impurities substantially perturb the
cluster matrix. It was found that a single metal atom drastically
modifies the properties of silicon clusters \cite{Hiura,Kumar}.
Therefore, it is interesting to explore the shape deformation and
the energy change induced by impurities. In this context rare gas
clusters are especially interesting. They are inert and could play
a role of the nanoscale chemical laboratory. A global optimization
is one of theoretical methods suitable for investigation of the
impurity-cluster interaction.

The most successful methods of global optimization for clusters
are genetic algorithms \cite{Xiao,Deaven,Wolf,Zeiri} and a
basin-hopping method
\cite{Wales,Doye,Leary,Naumkin1,Naumkin2,Miller,CCD}. The genetic
algorithm methods select a fraction of the initial population of
relaxed cluster structures as ``parents'' using their energies as
the criteria of fitness. The next generation of cluster structures
(``children'') is formed by ``mating'' these parents and the
process is repeated until the minima are found. The basin-hopping
approach of global optimization for clusters was developed by
Wales and Doye \cite{Wales}. This method uses the energy
minimization and the Monte Carlo simulation. The potential energy
is transformed to a new surface which has the form of a
multi-dimensional staircase. The steps correspond to the basins of
attraction surrounding energy minima \cite{Berry}. Each basin of
attraction contains the set of configurations which after geometry
optimization generate a particular minimum. The basin-hopping
method is very powerful and, for example, finds all known energy
minima for the Lennard--Jones clusters containing up to $110$
atoms \cite{Wales}. A similar method was used in the study of
protein folding \cite{Scheraga}. The Lennard--Jones nanoparticles
play a role of the test system for global optimization methods.
Molecule-doped rare gas clusters, in particular $H_{2}$ in
$Ar_{N}$ \cite{Zeiri}, $Cl_{2}$ in $Ar_{N}$ \cite{Naumkin1}, and
$NO$ in $Ar_{N}$ \cite{Naumkin2}, were also studied. In general,
the clusters with impurities are less studied by optimization
methods than homonuclear ones. A small perturbation, for example
one impurity atom, or ion, in a cluster may shed light on the
global minima formation. In addition, new morphologies in the set
of cluster structures may appear for some impurities. The methods
of global optimization produce the candidates for minima of the
free energy only close to $T=0$ K. At higher temperatures kinetics
effects are important and may change the order of stability of
cluster isomers.

In this work I present the results of structural optimization for model
heteronuclear clusters consisting of one impurity ion and of up to $N=79$ rare gas
atoms. The Lennard--Jones and Mason--Schamp potentials are used to describe
interactions. The basin-hopping method is applied for structural
optimization. The results show that an impurity ion is at the surface for a
small number of rare gas atoms. It is always trapped in a cluster for
$N\geq 8$ atoms. As in homonuclear Lennard--Jones clusters, several particularly stable
structures are found. In the following the computational method is described
in Sec. II. Results and discussion are presented in Sec. III. A summary and
conclusions are given in Sec. IV.

\section{Computational method}

\begin{figure}
\begin{center}
\includegraphics*[scale=.60]{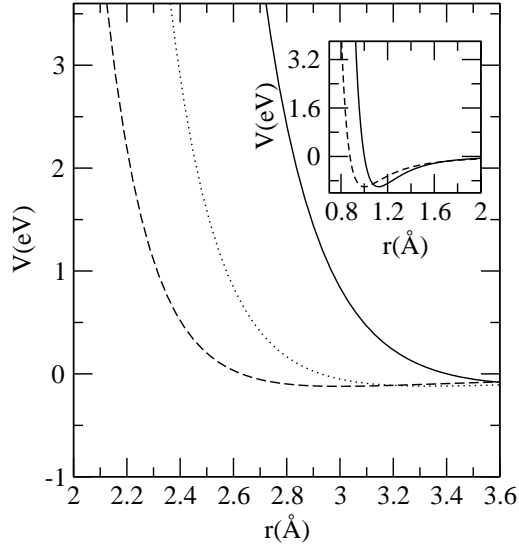}
\end{center}
\caption{The Mason--Schamp potential for $Cs^{+}$--$Xe $ (full
line), $Rb^{+}$--$Kr $ (dotted line), and $K^{+}$--$Ar $ (dashed
line). The inset shows a comparison of the Mason--Schamp potential
(dashed line) and the Lennard--Jones potential (full line) in
reduced units where $\epsilon=1$, $\sigma=1$.} \label{fig1}
\end{figure}

The pairwise additivity and the spherical symmetry of the interaction for
rare gas atoms are reasonably well described by the Lennard--Jones potential.
This potential is one of the most used models of interaction. It is given by
\begin{equation}
V(r) = 4 \epsilon \left [ \left (\frac {\sigma} {r}\right)^{12} - \left
(\frac {\sigma} {r}\right)^{6}\right ],                      \label{eq:1}
\end{equation}
where $\epsilon$ is the depth of the potential energy minimum, and
$2^{1/6}\sigma$ is the equilibrium pair separation \cite{Lennard}. The $\sim$
$r^{-6}$ is the leading term in the dispersion energy interaction for neutral
atoms. Mason and Schamp proposed the additional term $\sim$$r^{-4}$ to model
the interaction of an ion and a neutral rare gas atom \cite{Mason}. The
Mason--Schamp potential is given by
\begin{equation}
V(r) = \frac {\epsilon}{2} \left [(1 + \gamma) \left(\frac {\sigma}
{r}\right)^{12} - 4 \gamma \left(\frac {\sigma} {r}\right)^{6} - 3 (1 - \gamma)
\left(\frac {\sigma} {r}\right)^{4}\right ].                    \label{eq:2}
\end{equation}
The coefficients in this potential were determined from the ion
mobility measurements in rare gases and calculations within the
kinetic theory. Three parameters in algebraic form (\ref{eq:2})
specify the depth and position of the potential energy minimum,
and the relative strength of various terms. The Mason--Schamp
potential was used in the studies of impurities in rare gas
crystals \cite{Smirnov}. For example, the parameters of this
potential for the $Cs^{+}$--$Xe $ system are: $\epsilon = 0.106$
eV, $\sigma=3.88\AA$, and $\gamma = 0.2$. Figure \ref{fig1} shows
the Mason--Schamp and Lennard--Jones potentials. The Mason--Schamp
potential was taken here in its reduced form as a model of a class of the alkali-metal
ion and rare gas atom interaction. In the same sense the Lennard--Jones potential
represents interaction between rare gas atoms in investigations of the global
minima for homonuclear clusters \cite{Deaven,Wolf,Munro}. The Lennard-Jones and
Mason--Schamp potentials in the inset of Fig. 1 are drawn in reduced units where
$\epsilon$ and $\sigma$ are chosen as the units of energy and distance.
In the calculations the total
potential was modelled by a pairwise sum of potentials given by
Eq. (\ref{eq:1}) for each rare gas/rare gas pair of atoms, and by Eq. (\ref{eq:2}) for each rare gas/ion
pair of particles. For both potentials
reduced units for $\epsilon$ and $\sigma$ were employed.

I used a basin-hopping method and a modified program gmin of Wales and
collaborators \cite{CCD}. The constant temperature Monte Carlo method based
on the standard Metropolis scheme \cite{Binder} is applied. Energy
minimization is performed using the limited memory
Broyden--Fletcher--Goldfarb--Shanno (BFGS) algorithm for the
unconstrained minimization problem \cite{Liu}. For each cluster size five
series of runs were done starting from different configurations. Each run
consisted of $30000$ Monte Carlo steps at a constant reduced temperature of
$0.8$. Control runs for homonuclear Lennard--Jones clusters have been carried out,
and the same global minima were found as by Wales and Doye \cite{Wales}. For
sizes $N<75$, direct optimizations from random configurations were able to
produce known minima for homonuclear Lennard--Jones clusters within less than $10000$
Monte Carlo steps. For $N\geq 75$, additional effort was necessary, either
the increase of the number of steps, and/or seeding with configurations
derived from the best structures around a chosen cluster size.
For both homonuclear and heteronuclear
clusters the lowest-energy structures were found at different Monte Carlo steps in five series of runs. Sometimes,
different lowest-energy structures were found in these runs. Differences between various runs were especially
pronounced for $N\ge 75$. Similar type
of potentials are used for the atom-atom and atom-ion interactions and
it is plausible that the same procedure is adequate for unknown
energy minima of heteronuclear clusters. However, in heteronuclear systems the forces
are determined not only by distances between particles, but their chemical
identity is also important. In homonuclear clusters swapping the position of the particles
does not change structural properties of the aggregate. For the system and the method
presented in this work (where only one impurity particle, two similar potentials, and
the Monte Carlo minimization method are used) the optimization tasks for homonuclear and
heteronuclear clusters are analogous. In theoretical analysis of global optimization and
NP-hard problems, the case of heteronuclear clusters is more difficult \cite{Greenwood}.

\section{Results and discussion}

\begin{figure}
\begin{center}
\includegraphics*[scale=.65]{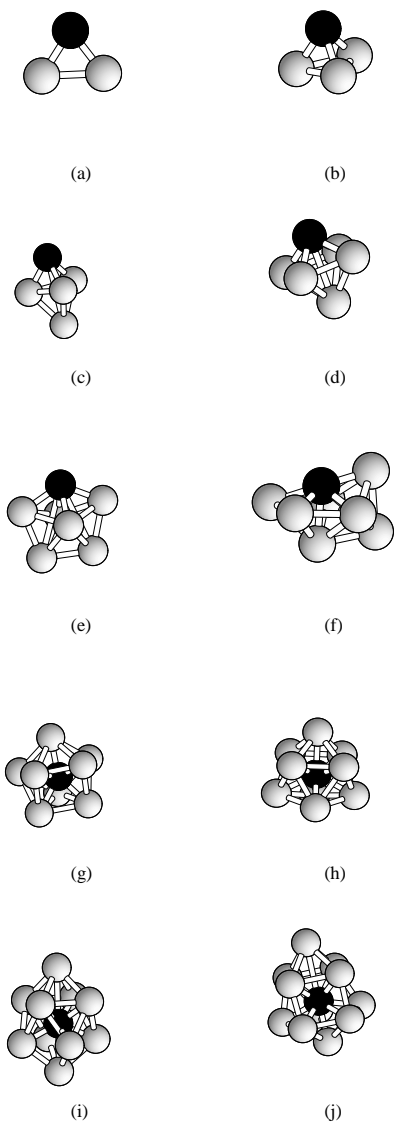}
\end{center}
\caption{The global minima configurations for one alkali-metal ion and:
(a) $2$, (b) $3$, (c) $4$, (d) $5$, (e) $6$, (f) $7$, (g) $8$, (h) $9$, (i) $10$,
(j) $11$ rare gas atoms. Light and dark balls correspond to the rare gas
atoms and the alkali-metal ion, respectively. When the number of rare gas
atoms is equal or greater than $8$, the impurity ion is always inside the
cluster.}
\label{fig2}
\end{figure}

\begin{figure}
\begin{center}
\includegraphics*[scale=.70]{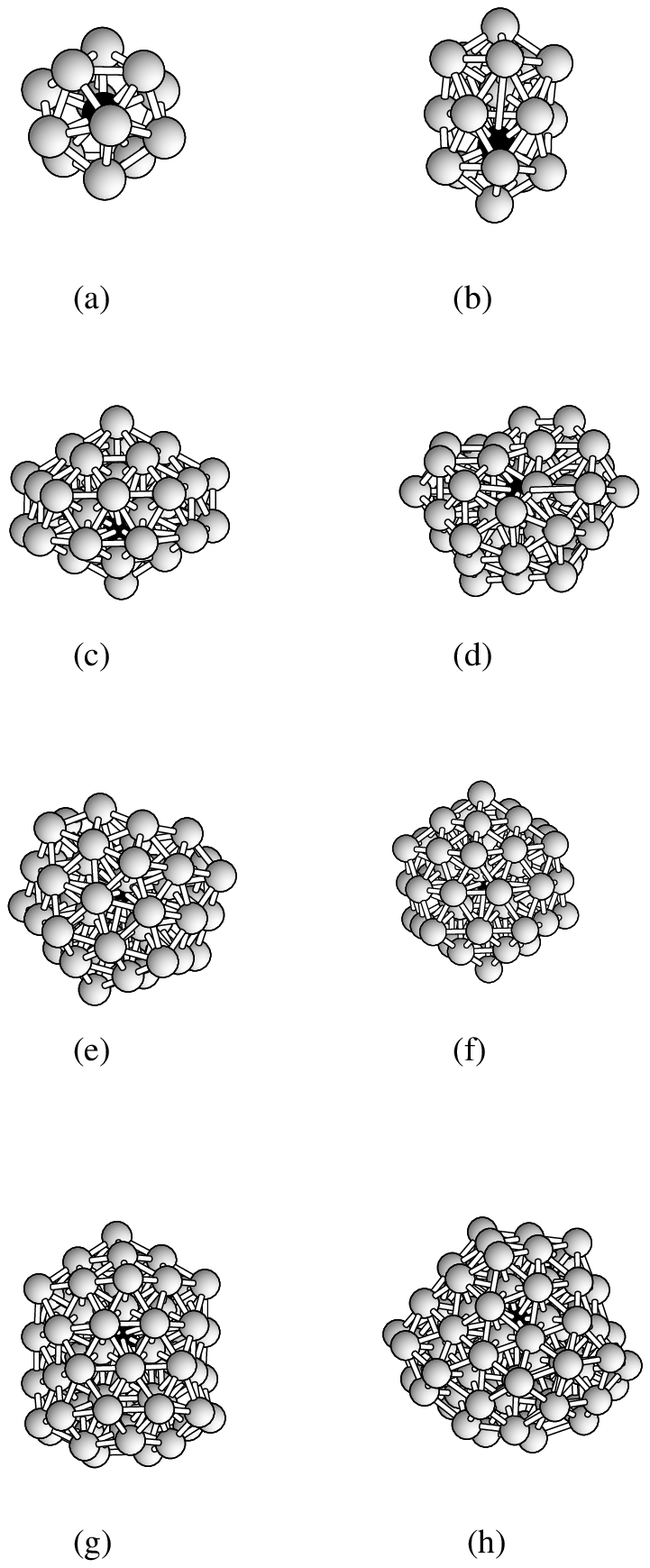}
\end{center}
\caption{The global minima configurations for $N+1$ equal to: (a) $13$,
(b) $19$, (c) $39$, (d) $46$, (e) $49$, (f) $55$, (g) $70$, (h) $79$ particles,
where $N$ is the number of rare gas atoms. These structures are the most
stable ones. Details as in Fig. \ref{fig2}.}
\label{fig3}
\end{figure}

Figures \ref{fig2} and \ref{fig3} show several configurations of an alkali-metal impurity in a
rare gas cluster.
This visualization is performed using the Rasmol package
\cite{Sayle,Bernstein}. It is found that all lowest-energy structures of
heteronuclear clusters are compact. The spacial distributions of particles does not
change much in comparison with homonuclear Lennard--Jones clusters. The
ion-atom distance is smaller than the atom-atom one. The lengths of the
bonds and the angles between them change slightly. The ratio of the numbers
of the atom-atom and ion-atom bonds increases with the cluster size.
Therefore, a difference between corresponding homonuclear and heteronuclear
clusters decreases when their size increases. The energy minima are presented in
Table \ref{table1}.
\begin{table}
\caption{Energy divided with $\epsilon$ for the calculated putative
lowest energy atomic configurations of $(N-1)$ rare gas atoms and one
alkali-metal ion.}
\label{table1}
\begin{tabular}{cccccccccc}
\hline
N&Energy/$\epsilon$&N&Energy/$\epsilon$&N &Energy/$\epsilon$&N&Energy/$\epsilon$&N&Energy/$\epsilon$\\
\hline
3&-3.000000     &19 & -74.117047  &35 &-157.959698   & 51 &-255.561479   &67 & -352.320332   \\
4 & -6.000000   &20 &-78.669480   & 36 & -164.131670 & 52 & -262.641668  & 68 & -358.620511  \\
5 & -9.170198   &21 & -83.223757  & 37 & -169.413923 & 53 & -269.723630  & 69 &-364.991246   \\
6 & -12.833944  &22 & -88.512937  & 38 &-175.688479  & 54 &-276.837164   & 70 & -372.140719  \\
7 & -16.609860  &23 &-94.747096   & 39 & -182.644656 & 55 & -283.982317  & 71 & -378.708743  \\
8 &-20.267149   &24 & -99.271061  & 40 & -187.934276 & 56 & -288.408595  & 72 &-384.001154   \\
9 & -24.603991  &25 & -104.437908 & 41 &-193.313553  & 57 &-293.124190   & 73 & -390.193626  \\
10 & -29.845315 &26 &-110.531823  & 42 &-199.171591  & 58 & -299.182087  & 74 & -396.341772  \\
11 &-34.559546  &27 & -115.202264 &43 & -205.377149  & 59 & -304.564977  & 75 &-401.715686   \\
12 & -38.823624 &28 & -120.217617 & 44 & -210.799981 & 60 &-310.725716   & 76 & -407.874621  \\
13 & -44.948660 &29 &-126.039309  & 45 &-217.099260  & 61 & -316.885236  & 77 & -414.039706  \\
14 &-48.507280  &30 & -130.940143 & 46 & -224.073078 & 62 & -322.259510  & 78 &-420.198993   \\
15 & -53.015985 &31 & -136.316374 & 47 & -229.513961 & 63 &-328.423273   & 79 & -427.364107  \\
16 & -57.557929 &32 &-141.606686  & 48 &-235.929062  & 64 & -334.583581  & 80 & -433.748290  \\
17 &-62.572600  &33 & -147.154918 & 49 & -242.915907 & 65 & -339.974002                     \\
18 & -67.609864 &34 & -152.348526 & 50 & -248.498878 & 66 &-346.146925                       \\
\hline
\end{tabular}
\end{table}

The situation where the alkali-metal ion is at the surface of a cluster is
shown in Fig. \ref{fig2}(a-f).
Figures \ref{fig2}(f) and \ref{fig2}(g) present configurations around
the transitional regime of the ion. In Fig. \ref{fig2}(f) the impurity ion is deeply
buried in the surface of the cluster, and lies closer to its centre than one
rare gas atom. The ion is in the interior of the cluster the first time for
eight rare gas atoms, as shown in Fig. \ref{fig2}(g). The impurity ion is also always
in the cluster for more than eight rare gas atoms (Figs. \ref{fig2}(h-j) and Fig. \ref{fig3}).
When the alkali-metal ion is solvated by rare gas atoms it is not always
situated in the centre of the cluster. For example, the impurity ion takes a
non-central position in clusters shown in Figs. \ref{fig3}(b-d).

\begin{figure}
\begin{center}
\includegraphics*[scale=.60]{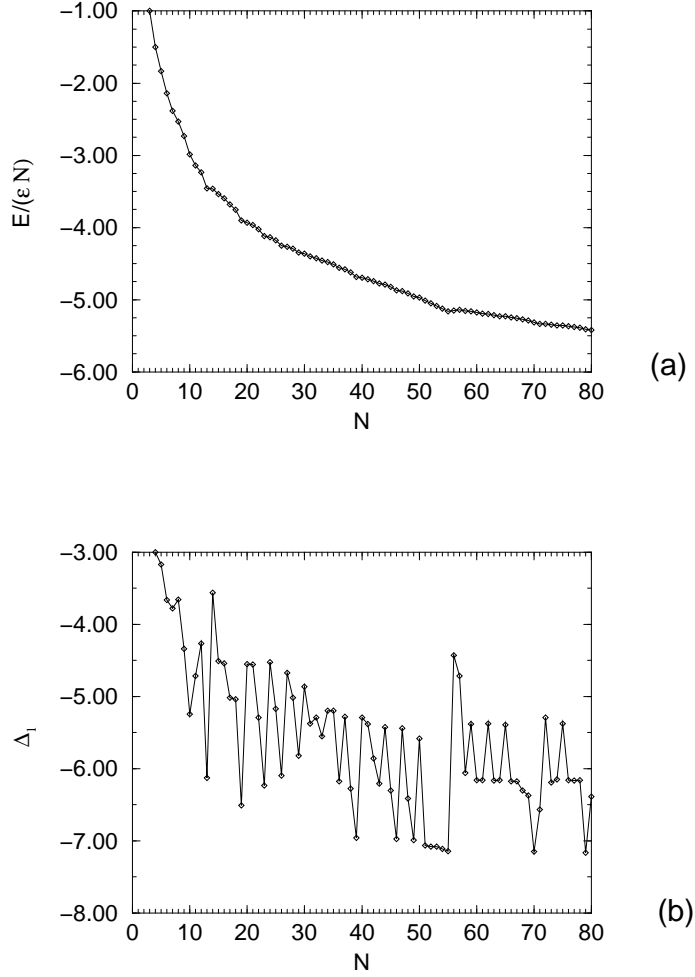}
\end{center}
\caption{(a) Energies per number of particles $N$, and (b) the first energy
difference, $\Delta_1 = (E(N)-E(N-1))/\epsilon$, for globally
optimized clusters plotted vs N.}
\label{fig4}
\end{figure}

The energies per particle are presented in Fig. \ref{fig4}(a). This quantity varies
rather monotonically. However, mild minima occur for certain sizes and show
particularly stable structures. This is further confirmed by the first
energy differences $\Delta _{1}=E(N)-E(N-1)$ shown in Fig. \ref{fig4}(b). Two minima
at $N=13$ and $N=55$ correspond to well known filled icosahedral shells. In
addition, minima occur for $N=19,39,46,49,70,79$. These most stable
structures of rare gas clusters with an alkali-metal ion are shown in
Fig. \ref{fig3}. Minima at $N=19,46,49,70,79$ also exist for Lennard--Jones clusters (for
example, see Fig. 4 in \cite{Wolf}). However, in homonuclear Lennard--Jones clusters
minimum exist for $N=38$.
Figure \ref{fig4}(b) shows that for heteronuclear clusters
corresponding minimum is $N=39$. Shapes of $N=38$ and $N=39$ (see Fig. \ref{fig3}(c))
heteronuclear clusters are similar to each other and to Lennard--Jones $N=39$. The
Lennard--Jones $N=38$ cluster is slightly more compact (see Fig. \ref{fig1} in
\cite{Miller}). A shape of the $N=39$ structure shown in Fig. \ref{fig3}(c) is closer to
the third lowest energy minimum of Lennard--Jones $N=38$, than to the first
and second one. In connection with the complex double-funnel energy
landscape of the  non-icosahedral $38$-atom Lennard--Jones cluster
\cite{Miller}, it is interesting to point out that this magic size in
heteronuclear nanoparticles is replaced by
$N=39$. Other non-icosahedral Lennard--Jones
global minima occur for $N=75$--$77$, and these clusters are Marks decahedra
\cite{Wales}. For clusters with the impurity ion, sizes $N=75$--$77$ are
decahedral in origin. As for icosahedral minima, these structures are
distorted in comparison with homonuclear Lennard--Jones clusters. The flat region in
Fig. \ref{fig4}(b) exist for $51\leq N\leq 55$. In these clusters the impurity ion is
situated in the central position and the magic icosahedral $N=55$ structure
builds up. For homonuclear Lennard--Jones clusters a such flat region is
$52\leq N\leq 55$ \cite{Wolf}, and it also corresponds
to the filling of the magic $N=55$ structure.

\section{Conclusions}

The global optimization based on a basin-hopping method is carried out to
study structural properties of model nanoparticles consisting of one
impurity alkali ion and a rare gas cluster. The catalog of energy minimum
candidates for these clusters containing up to $80$ particles is presented.
It is found that the impurity ion is lying in the surface of a nanoparticle
for a small number of atoms. An alkali-metal ion is solvated by rare gas
atoms when their number is greater or equal than eight. The predicted magic
numbers $13,19,46,49,55,70,79$ of heteronuclear clusters are the same as for
homonuclear Lennard--Jones ones.
The magic structure $N=39$ of heteronuclear clusters replaces
$N=38$ of homonuclear nanoparticles. Therefore a small perturbation, such as the
presence of the closest alkali-metal ion neighbour from the periodic table,
may change the list of magic sizes for rare gas clusters.
The system studied in this work is useful as a test example for theoretical
analysis of global optimization and NP-hard problems for heteronuclear
clusters \cite{Greenwood}.
The methods of
global optimization are helpful as a guide in low-temperature experimental
studies of nanoparticles with impurities.

\ack{ This work has been carried under the HR-MZT project 0119255
``Dynamical Properties and Spectroscopy of Surfaces and
Nanostructures''.}

\end{document}